\documentclass[a4paper,11pt]{article}
\pdfoutput=1 

\usepackage{dcolumn}
\usepackage{bm}
\usepackage{amsthm}
\usepackage{setspace}
\usepackage[normalem]{ulem}
\usepackage{relsize}
\usepackage{cancel}
\usepackage{verbatim}
\usepackage{enumitem}
\usepackage{mathtools}
\usepackage{empheq}
\usepackage{dsfont}
\usepackage{amsmath}
\usepackage{slashed}

\usepackage{graphicx}
\usepackage{caption}
\usepackage{subcaption}

\usepackage{jheppub}

\usepackage{amsthm}

\newcolumntype{I}[1]{>{\centering\arraybackslash$}m{#1}<{$}}

\title{${\cal N} =1$ Liouville SCFT in Four Dimensions}

\author{Tom Levy$^1$,}
\author{Yaron Oz$^1$,}
\author{Avia Raviv-Moshe$^1$}
\affiliation{$^1$ Raymond and Beverly Sackler School of Physics and Astronomy, Tel-Aviv University, 55 Haim Levanon street, Tel-Aviv, 69978, Israel}
\emailAdd{toml@mail.tau.ac.il}
\emailAdd{yaronoz@post.tau.ac.il}
\emailAdd{aviaravi@mail.tau.ac.il}

\abstract{We construct a four supercharges Liouville superconformal field theory in four dimensions. The Liouville superfield is chiral and its lowest
component is a log-correlated complex scalar whose real part carries a background charge. 
The action consists of a supersymmetric Paneitz operator, a background supersymmetric ${\cal Q}$-curvature charge and an exponential potential.
It localizes semiclassically on solutions that describe curved superspaces with a constant complex supersymmetric  ${\cal Q}$-curvature.
The theory is non-unitary with a continuous spectrum of scaling dimensions.
We study the dynamics on the supersymmetric 4-sphere, show that the classical background charge is not corrected quantum mechanically and
calculate the super-Weyl anomaly. We derive an integral form for the correlation functions of vertex operators. 
}

\begin{document}
\maketitle
\flushbottom

\section{Introduction}

The two-dimensional Liouville  conformal field theory (CFT) 
has been studied extensively since its
introduction by Polyakov in the context of non-critical string theory  \cite{Polyakov:1981rd}.
A generalization of Liouville CFT to higher dimensions that consists of a log-correlated real scalar field with a background $\mathcal{Q}$-curvature charge and an exponential Liouville potential has been studied in  \cite{Levy:2018bdc}.
This higher-dimensional
CFT  is part of the inertial range field theory of turbulence proposed in \cite{Oz:2017ihc,Oz:2018mdq}, where the Liouville field is a Nambu-Goldstone
boson, the background charge is related to the fluid intermittency and the Liouville potential is the local fluid energy dissipation.
The higher-dimensional Liouville CFT is non-unitary with a continuus spectrum of scaling dimensions. It localizes semiclassically on solutions that describe manifolds with a constant  $\mathcal{Q}$-curvature and has an analog  \cite{Levy:2018bdc,Furlan:2018jlv} of the  DOZZ \cite{Zamolodchikov:1995aa,Dorn:1992at}  three-point function formula.

Two-dimensional $\mathcal{N}=1$  Liouville superconformal field theory (SCFT) has been introduced in \cite{Polyakov:1981re} in the context of non-critical
superstring theory and the  $\mathcal{N}=2$ Liouville SCFT in  \cite{Ivanov:1983wp}. While sharing some of the features of the two-dimensional Liouville
CFT such as $\mathcal{N}=1$   \cite{Rashkov:1996np,Poghosian:1996dw} and $\mathcal{N}=2$ \cite{Fukuda:2002bv} analogs of the DOZZ three-point function formula,   there are also differences such as a non-renormalization 
of the background charge in the  the  $\mathcal{N}=2$ Liouville SCFT  \cite{Distler:1989nt,Mussardo:1988av}
and its duality to the two-dimensional fermionic black hole \cite{Hori:2001ax}.

The aim of this work is to construct and study an  $\mathcal{N}=1$  Liouville SCFT in four dimensions. 
This theory has four real supercharges as the  $\mathcal{N}=2$ Liouville SCFT  in two dimensions.
It is a higher derivative theory where the  Liouville superfield is chiral, its lowest
component is a log-correlated complex scalar and its highest component  
is a complex scalar that is dynamical rather than being auxiliary.
The action consists of a supersymmetric Paneitz operator \cite{Butter:2013lta}, a background supersymmetric ${\cal Q}$-curvature charge \cite{Butter:2013ura} and an exponential potential.
The theory is non-unitary with a continuous spectrum of scaling dimensions 
and it localizes semiclassically on solutions that describe curved superspaces with a constant complex supersymmetric  ${\cal Q}$-curvature.
We will consider the theory on the supersymmetric 4-sphere, show that the classical background charge is not corrected quantum mechanically,
calculate the super-Weyl anomaly and derive an integral form for the correlation functions of vertex operators.

The paper is organized as follows. In section \ref{sec:ClassicalModel} we will study the classical properties of the four-dimensional
Liouville SCFT.
We will construct the action by introducing a supersymmetric Panietz operator and a supersymmetric ${\cal Q}$-curvature
that corresponds to a background charge.
We will analyze the classical super-Weyl invariance of the theory
and derive the field equations.
Next we will consider the theory on the supersymmetric 4-sphere, analyze its noncompact R-symmetry and construct a solution to the classical
field equations.
In section \ref{sec:QuantumModel} we will study quantum aspects of the Liouville SCFT.
We will show that the classical value of the background charge is not corrected quantum mechanically, classify the primary operators of the theory
and calculate the super-Weyl anomaly coefficients. In section \ref{sec:CorrelationFunctions} we will consider the correlation functions of vertex operators. 
We will derive an integral expression for them by considering the relation to free fields and four-dimensional Coulomb gas integrals, discuss  
the semiclassical limit and the
light and heavy primary operators.
Section \ref{sec:Summary} is devoted to a summary and outlook.
In appendix \ref{app:SUGRANotations} we will give a brief summary of notations and conventions.
In appendix \ref{app:2DNequals1SLFT} we will calculate in the semiclassical approximation the three-point function of light vertex operators in two-dimensional 
$\mathcal{N}=1$  Liouville SCFT in agreement with the semiclassical limit of the exact correlation functions.

\section{Classical Liouville SCFT in Four Dimensions}
\label{sec:ClassicalModel}

In this section we will construct and study the classical aspects of ${\cal N} =1$ Liouville superconformal field theory in four dimensions.
We will use the notations of \cite{Wess:1992cp} that are briefly summarized in appendix \ref{app:SUGRANotations}. 
The supegravity multiplet consists of the
vielbein $e_m^a(x)$, the gravitino $\psi_m^\alpha(x), \bar{\psi}_{m\dot{\alpha}(x)}$ and the auxiliary fields  $b_a$ and $M$. 
The supergravity notations are in Lorenzian signature $(-,+,+,+)$.
For the Euclidean signature of the supersymmetric 4-sphere
we will follow the analytical continuation in \cite{Festuccia:2011ws}.

\subsection{The Classical Theory}
\label{sec:TheAction}

\subsubsection{Action}
  
The action of four-dimensional Liouville superconformal field theory in curved superspace reads:
\begin{equation}
\label{eq:SLCFTAction}
S_L(\Phi,\bar{\Phi}) = \frac{1}{8\pi^2}\int d^4x\,d^2\Theta\,\mathcal{E}\left(\Phi\hat{\mathcal{P}}\bar{\Phi}+4Q\hat{\mathcal{Q}}\Phi+16\pi^2\mu e^{3b\Phi}\right)+\text{h.c.}
\ ,
\end{equation}
where $h.c.$ refers to hermitian conjugation.
The Liouville superfield $\Phi$ is a chiral superfield that  consists of a 
complex Liouville field $\phi$, a Weyl fermion $\psi_\alpha$ and a complex scalar $F$:
\begin{equation}
\Phi = \phi(x) + \sqrt{2}\Theta^\alpha\psi_\alpha(x) + \Theta\Theta F(x) \ ,
\end{equation}
and the $\Theta$ variables carry local Lorentz indices. $\bar{\Phi}$ denotes the anti-chiral conjugate of $\Phi$.
The action \eqref{eq:SLCFTAction} describes a higher derivative theory where the Liouville field $\phi$ is log-correlated and
the complex scalar $F$ is dynamical rather than auxiliary.
$\mathcal{E}$ is the chiral density \eqref{EPS} and $d^4x\,d^2\Theta$ is the chiral curved superspace measure.

 The dimensionless complex parameters in  \eqref{eq:SLCFTAction} 
are the background charge $Q$, the cosmological constant  $\mu$ and $b$. Note that we can always transform to the case where the background charge $Q$ is real by redefining the phase of the Liouville superfield. 
In the following we will take $Q$ and $b$ to be real.
We will denote by $S_{C.G.}$ the action \eqref{eq:SLCFTAction} when  $\mu = 0$ which describes a four-dimensional  supersymmetric  Coulomb gas.

There are two objects in the action  \eqref{eq:SLCFTAction}, $\hat{\mathcal{P}}$ and $\hat{\mathcal{Q}}$,  that play an important role
in superconformal geometry.
The operator $\hat{\mathcal{P}}$ is a supersymmetric extension of the conformally covariant fourth-order differential  Paneitz operator 
\cite{Pan,Fradkin:1982xc,Fradkin:1981jc}:
\begin{equation}
\mathcal{P} = \Box^2 + \mathcal{D}_{a}\left(2\mathcal{R}^{ab}-\frac{2}{3}\mathcal{R}g^{ab}\right)\mathcal{D}_{b} \ ,
\label{P}
\end{equation}
where $\mathcal{D}_a$ is the covariant derivative and $\Box=g^{ab}\mathcal{D}_a\mathcal{D}_b$.
$\hat{\mathcal{P}}$  is a super-Weyl covariant differential operator and takes the form \cite{Butter:2013lta}:
\begin{equation}
\label{eq:SuperPaneitz}
\hat{\mathcal{P}} = -\frac{1}{64}\left(\bar{\mathcal{D}}^2-8R\right)\left(\mathcal{D}^2\bar{\mathcal{D}}^2+16\mathcal{D}^{\alpha}\left(G_{\alpha\dot{\alpha}}\bar{\mathcal{D}}^{\dot{\alpha}}\right)\right) \ ,
\end{equation}
where $\mathcal{D}_{\alpha}$ is the covariant superderivative and $R$ and $G$  are chiral and real superfields, respectively.
They contain the Ricci scalar and the Ricci tensor, respectively, and their lowest components are the auxiliary fields of the supergravity multiplet \eqref{AUX}.
$\hat{\mathcal{P}}$ acts on an anti-chiral superfield $\bar{\Phi}$ and generates a chiral superfield:
\begin{equation} \mathcal{D}_{\alpha}\bar{\Phi} = 0,~~~~ \bar{\mathcal{D}}_{\dot{\alpha}}\hat{\mathcal{P}}\bar{\Phi} = 0 \ . 
\end{equation}
The operator $\hat{\bar{\mathcal{P}}}$ conjugate to $\hat{\mathcal{P}}$  acts on a chiral superfield $\Phi$ and produces an anti-chiral one 
$\mathcal{D}_{\alpha}\hat{\bar{\mathcal{P}}}\Phi = 0$.

$\hat{\mathcal{Q}}$  is a supersymmetric extension of the $\mathcal{Q}$-curvature \cite{Q}:
\begin{equation}
\mathcal{Q} = -\frac{1}{6}\left(\Box \mathcal{R}+3\mathcal{R}^{ab}\mathcal{R}_{ab}-\mathcal{R}^2\right) \ ,
\label{Q}
\end{equation}
and is a super-Weyl covariant chiral superfield \cite{Butter:2013ura}:
\begin{equation}
\label{eq:SuperQCurv}
\hat{\mathcal{Q}} = - \frac{1}{2} \left( \bar{\mathcal{D}}^2-8R\right)\left( G^a G_a + 2R \bar{R} -\frac{1}{8}\mathcal{D}^2 R \right) \ ,
\end{equation}
where $G_a \equiv  = \frac{1}{2}\sigma_a^{\alpha\dot{\alpha}}G_{\alpha\dot{\alpha}} $.

The Coulomb gas action \eqref{eq:SLCFTAction}  written in terms of   $\hat{\mathcal{P}}$ and
$\hat{\mathcal{Q}}$  is an extension to curved superspace of the four-dimensional Coulomb gas action 
written in terms of  $\mathcal{P}$ and
$\mathcal{Q}$  \cite{Levy:2018bdc}.
When setting the gravitino and the auxiliary fields of the supergravity multiplet to zero and writing the supersymmetric Coulomb gas action $S_{C.G.}$ in components, the action for the real part of the lowest component of the Liouville superfield $\mathrm{Re}(\phi)$ reduces to the standard bosonic
Coulomb gas action  \cite{Levy:2018bdc}.

\subsubsection{Super-Weyl Invariance}

Super-Weyl transformations \cite{Howe:1978km} are supersymmetric generalizations of the bosonic Weyl transformations
that  are parametrized by  a   chiral superfield $\sigma$ and  its anti-chiral conjugate $\bar{\sigma}$. 
The lowest component of $\sigma$ is a complex scalar $\frac{1}{2}(\omega+i\alpha)$, where $\omega(x)$
is the transformation parameter of  the bosonic Weyl transformation $\delta_{\omega} g_{mn} =2\omega g_{mn}$, while  $\alpha(x)$ 
corresponds to a chiral transformation of the gravitino \cite{Howe:1978km}.

Infinitesimally we have:
\begin{equation}
\label{eq:SuperWeylVielbein}
\begin{aligned}
\delta_{\sigma} E_{M}^{\quad\! a} &= (\sigma+\bar{\sigma})E_{M}^{\quad\! a} \ , \\
\delta_{\sigma} E_{M}^{\quad\! \alpha} &= (2\bar{\sigma}-\sigma)E_{M}^{\quad\! \alpha}+\frac{i}{2}E_{M}^{\quad\! a}\bar{\sigma}_{a}^{\dot{\alpha}\alpha}\bar{\mathcal{D}}_{\dot{\alpha}}\bar{\sigma} \ ,\\
\delta_{\sigma} E_{M}^{\quad\! \dot{\alpha}} &= (2\sigma-\bar{\sigma})E_{M}^{\quad\! \dot{\alpha}}+\frac{i}{2}E_{M}^{\quad\! a}\bar{\sigma}_{a}^{\dot{\alpha}\alpha}\mathcal{D}_{\alpha}\sigma \ ,
\end{aligned}
\end{equation}
where $E_M$ is the super vielbein.
Transformations \eqref{eq:SuperWeylVielbein} imply:
\begin{equation}
\label{eq:SuperWeyl}
\begin{aligned}
\delta_{\sigma} R &= -2(2\sigma-\bar{\sigma})R-\frac{1}{4}\bar{\mathcal{D}}^2 \bar{\sigma} \ , \\
\delta_{\sigma} G_{a} &= -(\sigma+\bar{\sigma})G_{a} +i\mathcal{D}_{a}(\bar{\sigma}-\sigma) \ .
\end{aligned}
\end{equation}

Using \eqref{eq:SuperWeylVielbein} and \eqref{eq:SuperWeyl} one can verify that the supersymmetric Paneitz operator $\hat{\mathcal{P}}$  \eqref{eq:SuperPaneitz}
and supersymmetric $\mathcal{Q}$-curvature $\hat{\mathcal{Q}}$ \eqref{eq:SuperQCurv}, transform covariantly under super-Weyl transformations:
\begin{equation}
\label{eq:SuperPaneitzQcurvTrans}
 \delta_{\sigma}\hat{\mathcal{P}} = -6\sigma \hat{\mathcal{P}}, \quad  \delta_{\sigma}\hat{\mathcal{Q}} = -6\sigma\hat{\mathcal{Q}}+\hat{\mathcal{P}}\bar{\sigma} \ ,
\end{equation}
where in our notations
they carry a super-Weyl weight six. 
\eqref{eq:SuperPaneitzQcurvTrans} are the super-Weyl generalizations of the Weyl transformations of the Paneitz operator \eqref{P}  and
the ${\cal Q}$-curvature \eqref{Q}:
\begin{equation}
\delta_{\omega}\mathcal{P} = -4\omega\mathcal{P},~~~~
\delta_{\omega}\mathcal{Q} = -4\omega\mathcal{Q}+\mathcal{P}\omega \ .
\end{equation}

The chiral measure carries weight minus six. 
The finite super-Weyl transformation of the super-Panietz operator and the super-$\mathcal{Q}$-curvature read:
\begin{equation}
\label{eq:SuperQCurvWeylTrans}
\hat{\mathcal{P}} \rightarrow  e^{-6\sigma}\hat{\mathcal{P}},~~~~
  \hat{\mathcal{Q}} \rightarrow e^{-6\sigma}\left(\hat{\mathcal{Q}}+\hat{\mathcal{P}}\bar{\sigma}\right) \ .
\end{equation}
The Liouville superfield $\Phi$ transforms under a super-Weyl transformation as:
\begin{equation}
\label{eq:PhiWeylTrans}
\Phi \rightarrow \Phi  -2Q\sigma,\quad \bar{\Phi} \rightarrow \bar{\Phi}  -2Q\bar{\sigma} \ .
\end{equation}

Using  \eqref{eq:SuperQCurvWeylTrans} and \eqref{eq:PhiWeylTrans} we get that the supersymmetric  Liouville action \eqref{eq:SLCFTAction} is 
classically invariant under  a super-Weyl  transformation:
\begin{equation}
\label{eq:ActionWeylTrans}
S_{L}(\Phi, \bar{\Phi}) \rightarrow  S_{L}(\Phi, \bar{\Phi}) - S_{C.G.}(2Q\sigma, 2Q\bar{\sigma}) \ ,
\end{equation}
with the background charge:
\begin{equation}
Q=\frac{1}{b} \ .
\label{Qcharge}
\end{equation}
Note, that the exponential potential is super-Weyl invariant by itself.

In order to derive \eqref{eq:ActionWeylTrans} we used: \footnote{For $\Phi_1 = \Phi_2$ the component form of the real functional resulting from \eqref{eq:SuperPaneitzForm} was written in \cite{Fradkin:1985am} up to terms involving the gravitino.}
\begin{equation}
\label{eq:SuperPaneitzForm}
\begin{aligned}
\int d^4x\, d^2\Theta\, 2\mathcal{E}\, \Phi_1\hat{\mathcal{P}}\bar{\Phi}_2 &= \int d^4x \, d^2\bar{\Theta}\, 2\bar{\mathcal{E}}\, \bar{\Phi}_2\hat{\bar{\mathcal{P}}}\Phi_1\\
&= \int d^4x\, d^2\theta\, d^2\bar{\theta}\, E\, \left(\frac{1}{16}\mathcal{D}^2\Phi_1\bar{\mathcal{D}}^2\bar{\Phi}_2-\mathcal{D}^{\alpha}\Phi_1 G_{\alpha\dot{\alpha}}\mathcal{D}^{\dot{\alpha}}\bar{\Phi}_2\right) \ ,
\end{aligned}
\end{equation}
where $\Phi_1,\Phi_2$ are chiral superfields, $E$ is the full superspace integration measure and $\theta$ carry Einstein indices. 
We see from \eqref{eq:SuperPaneitzForm} that $\hat{\mathcal{P}}$ and $\hat{\bar{\mathcal{P}}}$
are related to a Hermitian form, and that the kinetic term in \eqref{eq:SLCFTAction} can be written in a K\"{a}hler form.
The Hermitian form is super-Weyl invariant when $\Phi_1,\Phi_2$ carry zero weight.

The super-Weyl transformations $\eqref{eq:SuperPaneitzQcurvTrans}$ imply
that  the integral of the supersymmetric $\mathcal{Q}$-curvature over chiral superspace:
\begin{equation}
\int d^{4}x\,d^2\Theta\,2\mathcal{E} \hat{\mathcal{Q}} \ ,
\end{equation} 
 is super-Weyl invariant.
 This is analgous to the integral of the ${\cal Q}$-curvature \eqref{Q} which is an invariant of the conformal geometry.
One can construct the object \cite{Ferrara:1977mv,Bonora:1984pn,Buchbinder:1988tj}:
\begin{equation}
\label{eq:SuperGaussBonnet}
\int d^{4}x\,d^2\Theta\,2\mathcal{E}\left(\hat{\mathcal{Q}}+2W^{\alpha\beta\gamma}W_{\alpha\beta\gamma}\right) = 4\pi^2\left(\chi+ip\right) \ ,
\end{equation}
where $W_{\alpha\beta\gamma} $ is a chiral superfield symmetric in all its indices that contains the Weyl tensor.
It transforms under super-Weyl transformation as:
\begin{equation}
\delta_{\sigma}W_{\alpha\beta\gamma} = -3\sigma W_{\alpha\beta\gamma} \ .
\end{equation} 
The integral \eqref{eq:SuperGaussBonnet} is super-Weyl invariant but  it is not a supermanifold topological invariant of the curved superspace.
However, when setting to zero the gravitino and the auxiliary fields it is a topological invariant 
of the curved space, where 
$\chi$ and $p$ are  the Euler characteristic and first Pontryagin invariant, respectively.

\subsubsection{Field Equations}

 The field equations derived from the  action \eqref{eq:SLCFTAction} read:
\begin{equation}
\hat{\mathcal{P}}\bar{\Phi} + 2Q\hat{\mathcal{Q}} = -24\pi^2\mu b e^{3b\Phi}, \qquad   \hat{\bar{\mathcal{P}}}\Phi + 2Q\hat{\bar{\mathcal{Q}}} = -24\pi^2\bar{\mu} b e^{3b\bar{\Phi}} \ .
\end{equation}
Using  \eqref{eq:SuperQCurvWeylTrans} one sees that a solution to these field equations 
can be viewed as super-Weyl transformation parameters $\sigma = \frac{b}{2}\Phi$ and $\bar{\sigma} = \frac{b}{2}\bar{\Phi}$ that transform
 the background curved superspace into one with a constant complex super-$\mathcal{Q}$-curvature:
 \begin{equation}
\hat{\mathcal{Q}} = -12\pi^2\mu b^2,~~~~ \hat{\bar{\mathcal{Q}}} = -12\pi^2\bar{\mu} b^2 \ .
\end{equation}
Note, in comparison, that the field equations in the non-supersymmetric Liouville field theory studied in \cite{Levy:2018bdc}
have a real positive cosmological constant parameter $\mu$ and their solutions
can be viewed as Weyl factors that transform the background curved space into a constant negative $\mathcal{Q}$-curvature one.

 \subsection{Liouville SCFT on $S^4$}
\label{sec:LiouvilleSCFTonSphere}

\subsubsection{Background Charge}
\label{sec:BagroundCharge}

Consider a superconformally flat supermanifold, i.e.\ one that satisfies $W_{\alpha\beta\gamma} = 0$, with the topology of supersymmetric $S^4$. For a constant shift of the chiral superfield by a complex number $\phi_0$ we get: 
\begin{equation}
\label{eq:ActionShiftBC}
S_{C.G.}(\Phi+\phi_0)=S_{C.G.}(\Phi)+4Q\mathrm{Re}(\phi_0) \ .
\end{equation}
In deriving  \eqref{eq:ActionShiftBC} we used \eqref{eq:SuperGaussBonnet}
with $\chi = 2$ and $p=0$, which are the Euler characteristic and first Pontryagin invariant of $S^4$.  
One can check these values by plugging in \eqref{eq:ActionShiftBC}
the background auxiliary fields of the supergravity multiplet for a  rigid supersymmteric theory on $S^4$ \cite{Festuccia:2011ws}.

Since the $S^4$ supermanifold is superconformally equivalent to supersymmetric flat space, we can preform a (singular) super-Weyl transformation and work with flat space super vielbein. Following equation \eqref{eq:PhiWeylTrans}, assuming  a regular solution on the sphere $S^4$, this singular transformation results in a boundary condition for the Liouville superfield $\Phi$:
\begin{equation}
\label{eq:BoundaryCond}
 \Phi= -2Q\log\left(|x|\right)+O(1), \qquad  |x|\to\infty \ .
 \end{equation}
Writing \eqref{eq:BoundaryCond}  in component form we get:
\begin{equation}
\mathrm{Re}(\phi) = -2Q\log\left(|x|\right)+O(1), \qquad  |x|\to\infty \ ,
\label{REQ}
\end{equation}
 while all the other component fields $\mathrm{Im}(\phi),\psi,F$ approach a finite limit. 
 As a result of \eqref{REQ} the real part of the lowest component  $\phi$ of the Liouville 
 superfield  $\mathrm{Re}(\phi)$ has a background charge $Q$.

\subsubsection{Action and $R$-Symmetry}

Using the super-Weyl mapping of the supersymmetric sphere $S^4$ to a supersymmetric flat space background
the action \eqref{eq:SLCFTAction} reads:
\begin{equation}
\label{eq:FlatSuperspaceAction}
S_L = \frac{1}{128\pi^2}\int d^4x\, d^2\theta\, d^2\bar{\theta}\, D^2\Phi\bar{D}^2\bar{\Phi} +\left[ \int d^4x\, d^2\theta\, \mu e^{3b\Phi} +\text{h.c.}\right],
\end{equation}
where the Liouville superfield reads:
\begin{equation}
\Phi = \phi (x_+) +\sqrt{2} \theta \psi(x_+) +\theta \theta F(x_+) \ ,
\end{equation}
and $x_+^m =x^m + i\theta\sigma^m\bar\theta$ is the chiral coordinate, $\bar{D}_{\dot{\alpha}}x_+^m=0$. The conjugate anti-chiral coordinate is denoted by $x_-^m =x^m - i\theta\sigma^m\bar\theta$.
In components \eqref{eq:FlatSuperspaceAction} takes the form:
\begin{equation}
\label{eq:FlatComponentAction}
S_L = \frac{1}{8\pi^2}\int d^4x\left(\phi^{\dagger}\Box^2\phi+\bar{\psi}\Box\bar{\sigma}^a\partial_{a}\psi+F^{\dagger}\left(-\Box\right) F+ \left[24\pi^2\mu b\left(F-\frac{3}{2}b \psi\psi\right)e^{3b\phi}+ \text{h.c.}\right] \right) \ .
\end{equation}

Liouville SCFT on the sphere is invariant under an R-symmetry.
This is not the standard 
$U(1)_R$ symmetry:
\begin{equation}
\Phi'\left(e^{i\alpha}\theta, x_+\right) = e^{2in\alpha}\Phi\left(\theta,x_+\right),~~~~ 
 \bar{\Phi}'\left(e^{-i\alpha}\bar{\theta},x_-\right) = e^{-2in\alpha}\bar{\Phi}\left(\bar{\theta},x_-\right) \ ,
\end{equation}
which is a symmetry of the higher-derivative kinetic term in the action \eqref{eq:FlatSuperspaceAction}
but is broken both by the interaction term and by the boundary condition \eqref{eq:BoundaryCond}.
Rather, under the R-symmetry
the Liouville superfield transforms in an affine representation:
\begin{equation}
\label{eq:SuperspaceRSymm}
\begin{aligned}
\Phi'\left(e^{i\alpha}\theta,x_+\right) &= \Phi(\theta,x_+)+\frac{2i}{3b}\alpha,~~~~
\bar{\Phi}'\left(e^{-i\alpha}\bar{\theta},x_-\right) &= \bar{\Phi}\left(\bar{\theta},x_-\right)-\frac{2i}{3b}\alpha \ .
 \end{aligned}
 \end{equation}
  In component form it reads:
 \begin{equation}
 \label{eq:ComponentRSymm}
 \phi \to \phi + \frac{2i}{3b}\alpha,\quad \psi\to e^{-i\alpha}\psi, \quad F\to e^{-2i\alpha}F \ .
 \end{equation}
The R-symmetry group is noncompact and isomorphic to $\mathbb{R}$, thus
the R-charges are not quantized.

\subsubsection{Classical Solution}

The field equations read:
\begin{equation}
\label{eq:SuperfieldEOM}
-\frac{1}{64}D^2\bar{D}^2D^2\Phi  = -24\pi^2\bar{\mu} b e^{3b\bar{\Phi}} \ ,
\end{equation}
and in component form:
 \begin{equation}
   \label{eq:ComponentEOM}
   \begin{aligned}
   &\begin{aligned}
   \Box^2\phi &= -72\pi^2\bar{\mu} b^2\left(F^{\dagger}-\frac{3}{2}b\bar{\psi}\bar{\psi}\right)e^{3b\phi^{\dagger}}\ ,\\
   \Box\slashed{\partial}\psi &= 72\pi^2\bar{\mu} b^2 \bar{\psi}e^{3b\phi^{\dagger}}\ ,\\
   -\Box F &= -24\pi^2\bar{\mu} b e^{3b\phi^{\dagger}} \ .
   \end{aligned} 
   \end{aligned}
   \end{equation}
   The field equations are supplemented by the boundary conditions:
  \begin{equation}
    \phi = -\frac{2}{b}\log\left(|x|\right)+O(1), \qquad |x| \to \infty \ ,
   \end{equation} 
   where we used the classical value $Q=\frac{1}{b}$ of the background charge.

A solution to the equations of motion \eqref{eq:SuperfieldEOM} for the Lioville superfield is given by:
\begin{equation}
\label{eq:SuperSol}
\Phi(\theta,x_+) = -\frac{1}{b}\log\left(\left|\pi^2b^2\mu\right|^{2/3}|x_+|^2+1\right)+\frac{1}{3b}\log{\frac{2}{3}}-\theta\theta \frac{2\bar{\mu}}{b|\mu|}\frac{\left|\pi^2b^2\mu\right|^{1/3}}{\left|\pi^2b^2\mu\right|^{2/3}|x_+|^2+1} \ .
\end{equation}
Note,  that the fermionic part of the solution \eqref{eq:SuperSol} is zero. The solution \eqref{eq:SuperSol} is obtained by an analytical
 continuation to complex radii of the super-Weyl factor  of the supersymmetric sphere $S^4$ defined in  \cite{Festuccia:2011ws}. 
 
An analogous method applied to the  non-supersymmetric Liouville theory allows to write a solution for a constant positive $\mathcal{Q}$-curvature version of the equations. However, the solution cannot be analytically continued to a real function that solves the physically relevant constant negative $\mathcal{Q}$-curvature equations. In the supersymmetric case, the complex nature of the component fields allows for an analytic continuation to exist for all values of the complex parameter $\mu, \bar{\mu}$.

\section{Quantum Liouville SCFT  in Four Dimensions}
\label{sec:QuantumModel}

In this section we study quantum aspects of the Liouville SCFT. We classify the primary operators, their scaling dimensions and
R-charges, show that the background charge is not corrected quantum mechanically and calculate supersymmetric Weyl anomaly coefficients.

\subsection{Primary Operators and Background Charge Non-Renormalization}
\label{sec:NonRenormalizationAndPrimaryOperators}

The supersymmetric  four-dimensional  Coulomb gas theory, \eqref{eq:FlatComponentAction} when $\mu=0$ is a free SCFT 
that consists of 
an ordinary four-dimensional Coulomb gas field $\mathrm{Re}(\phi)$  with a background charge $Q$ \cite{Levy:2018bdc}, a conformal 
four-derivative real scalar $\mathrm{Im}(\phi)$, a conformal three-derivative Weyl fermion $\psi$ \cite{Bergshoeff:1980is} and its conjugate \cite{Fradkin:1985am} and a conformal two-derivative complex scalar $F$. 
The free field correlations functions read:
\begin{equation}
\begin{aligned}
\label{eq:FreePropagators}
\left<\phi(x)\phi^{\dagger}(y)\right>_{C.G.} &= -\log(|x-y|)\ ,\\
\left<F(x)F^{\dagger}(y)\right>_{C.G.} &= -\Box\left<\phi(x)\phi^{\dagger}(y)\right>_{C.G.}= \frac{2}{|x-y|^2}\ ,\\
\left<\psi_{\alpha}(x)\bar{\psi}_{\dot{\beta}}(y)\right>_{C.G.} &= -\sigma_{\alpha\dot{\beta}}^{m}\partial_{m}\left<\phi(x)\phi^{\dagger}(y)\right>_{C.G.}= \sigma_{\alpha\dot{\beta}}^{m}\frac{x_m}{|x-y|^2} \ .
\end{aligned}
\end{equation}

Consider the vertex operator:
\begin{equation}
\label{eq:BosonicVertexOp}
V_{\alpha\tilde{\alpha}} = e^{2\alpha\phi+2\tilde{\alpha}\phi^{\dagger}} = e^{2(\alpha+\tilde{\alpha})\mathrm{Re}(\phi) + 2i(\alpha-\tilde{\alpha})\mathrm{Im}(\phi)},
\end{equation}
where $\alpha, \tilde{\alpha}$ are two independent complex numbers. When $\alpha = \tilde{\alpha}$ we denote $V_{\alpha}=V_{\alpha\tilde{\alpha}}$. The vertex operators have scaling dimensions:
\begin{equation}
\label{eq:DimBosonicVertexOp}
\Delta_{\alpha\bar{\alpha}} = -4\alpha\tilde{\alpha}+2Q(\alpha+\tilde{\alpha}) \ ,
\end{equation}
and R-charges  \eqref{eq:ComponentRSymm}:
\begin{equation}
\label{eq:U1chargeOfVaa}
r_{\alpha\tilde{\alpha}} = \frac{4}{3b}\left(\alpha-\tilde{\alpha}\right) \ .
\end{equation}
Equations \eqref{eq:DimBosonicVertexOp}, \eqref{eq:U1chargeOfVaa} are invariant under 
\begin{equation}
\alpha \to \frac{1}{b}-\tilde{\alpha},\quad \tilde{\alpha} \to \frac{1}{b}-\alpha \ ,
\end{equation}
which suggests that $V_{\alpha,\tilde{\alpha}}$ and $V_{1/b-\tilde{\alpha},1/b-\alpha}$ correspond to the same quantum operator, up to normalization:
\begin{equation}
\label{eq:RefCoef}
V_{1/b-\tilde{\alpha},1/b-\alpha}=R(\alpha,\tilde{\alpha})V_{\alpha,\tilde{\alpha}} \ .
\end{equation}
As in two dimensions we will call  $R(\alpha,\tilde{\alpha})$ a reflection coefficient.

Requiring the Liouville interaction to be a marginal operator implies:
\begin{equation}
\Delta(e^{3b\phi}) = \Delta_{\frac{3}{2}b,0} = 3 \ ,
\end{equation}
and using \eqref{eq:DimBosonicVertexOp} we get the quantum value of the background charge:
\begin{equation}
\label{eq:NonRenormQ}
Q = \frac{1}{b} \ .
\end{equation}
We see that the classical value of the background charge \eqref{Qcharge} is not corrected quantum mechanically.
Note that $\mathcal{N}=2$ Liouville SCFT in two dimensions \cite{Distler:1989nt,Mussardo:1988av} 
exhibits a similar 
non-renormalization  of the background charge.

The dimensions and R-charges of the fermions and the complex scalar read:
\begin{equation}
 \Delta_{\psi}=\Delta_{\bar{\psi}}=\frac{1}{2}~~~r_{\psi}=-r_{\bar{\psi}}=-1,~~~~\Delta_{F}=\Delta_{F^\dagger}=1~~~
r_{F}=-r_{F^{\dagger}}=-2 \ ,
 \end{equation}
 and one constructs primary operators by multiplying the vertex operator \eqref{eq:BosonicVertexOp} with an integer power of these fields. Consider
 the operator $\mathcal{O}$ with $m_\psi,\, m_{\bar{\psi}}$ insertions of the fermionic fields and $m_{F},\, m_{F^{\dagger}}$ insertions of 
 the complex scalar. Its dimension and R-charge read:
 \begin{align}
&\Delta_{\mathcal{O}} =  -4\alpha\tilde{\alpha}+2Q\left(\alpha+\tilde{\alpha}\right)+\frac{1}{2}\left(m_{\psi}+m_{\bar{\psi}}\right)+m_{F}+m_{F^{\dagger}},\nonumber\\
& r_{\mathcal{O}} =\frac{4}{3b}\left(\alpha-\tilde{\alpha}\right)+m_{\bar{\psi}}-m_{\psi}+2\left(m_{F^{\dagger}}-m_F\right).
\end{align}
In particular, the scaling dimension and R-charge of the chiral primary operator:
\begin{equation}
\mathcal{O}_\alpha =  e^{\alpha\Phi},~~~~
\Delta_{\mathcal{O}_\alpha} = Q\alpha,~~~r_{_{\mathcal{O}_{\alpha}}} = \frac{2\alpha}{3b} \ ,
\end{equation}
satisfy $r_{_{\mathcal{O}_{\alpha}}}  = \frac{2}{3}\Delta_{\mathcal{O}_\alpha}$ as expected.

\subsection{Super-Weyl Anomalies}
\label{sec:SuperWeylAnom}

The supertrace $\mathcal{T}$  is the supersymmetric version of the trace of the stress-energy tensor and
 is a chiral superfield defined as the variation of the action under a super-Weyl transformation.
The supertrace vanishes on-shell  when the classical action is super-Weyl invariant. 
The quantum expectation value of the supertrace is the variation of the quantum action $W = - \log Z$ under a super-Weyl transformation:
\begin{equation}
\label{eq:SuperTracDef}
 \left<\mathcal{T}\right> = \frac{\delta W}{\delta \sigma} \ .
 \end{equation} 
The super-Weyl symmetry of the classical theory is in general anomalous at the quantum level and the
expectation value for the supertrace \eqref{eq:SuperTracDef} is nonzero on a curved superspace background. 

The structure of the super-Weyl anomaly reads \cite{McArthur:1983fk, Bonora:1984pn, Buchbinder:1986im}:
\begin{equation}
\label{eq:SuperTraceCoeff}
\left<\mathcal{T}\right> = \frac{c}{2\pi^2} W^{\alpha\beta\gamma}W_{\alpha\beta\gamma} - \frac{a}{8\pi^2}\mathcal{G} + h\left(\bar{\mathcal{D}}^2-8R\right)\mathcal{D}^2R \ .
\end{equation}
$\mathcal{G}$ is a topological density \cite{Ferrara:1977mv}:
\begin{equation}
\mathcal{G} = 4W^{\alpha\beta\gamma}W_{\alpha\beta\gamma}- \left(\bar{\mathcal{D}}^2-8R\right)\left(G^aG_a+2R\bar{R}\right) \ ,
\end{equation}
and its chiral superspace integral is related to the Euler characteristic and first Pontryagin invariant by
 \eqref{eq:SuperGaussBonnet}. 
 The supertrace $\eqref{eq:SuperTraceCoeff}$ can be related to the Weyl anomaly
by setting the gravitino and the auxiliary fields of the supergravity multiplet to zero.
The anomaly coefficients $a$ and $c$ correspond to the A-type and B-type Weyl anomalies \cite{Deser:1993yx,Anselmi:1997am}. The coefficient $h$ is 
regularization dependent and can be changed by adding a finite local counterterm to the quantum action.

The Weyl  anomaly coefficients of the Liouville SCFT do not depend on the interaction terms
and are the same as those of the supersymmetric Coulomb gas theory by the following reasoning. One
can calculate the anomaly coefficients by considering correlation functions of the stress-energy tensor in a state of our choice. The interaction terms in the action \eqref{eq:SLCFTAction} are invariant under an arbitrary variation of the
metric, hence the stress-energy tensors of  Liouville SCFT and supersymmetric Coulomb gas theory are equal. By choosing to the
evaluate correlation functions of the stress-energy tensor in a state in which $\mathrm{Re}(\phi)\ll 0$ we can turn off the interaction terms and reduce the calculation of the anomaly coefficients to the supersymmetric Coulomb gas theory.
The latter calculation  is straightforward since we work with free fields and is simply the sum of the Weyl anomalies 
of the component fields listed in the table   \cite{Fradkin:1981jc,Fradkin:1985am,Levy:2018bdc,Duff:1993wm}:

\begin{center}
  \begin{tabular}{| c | c | c | }
    \hline
    $\mathrm{Re}(\phi)\;\;\;\mathrm{theory}$ & $ a=-\frac{7}{90}-Q^2$ & $c=-\frac{1}{15}$ \\ \hline
    $\mathrm{Im}(\phi) \;\;\;\mathrm{theory}$ & $a=-\frac{7}{90}$ & $c=-\frac{1}{15}$ \\ \hline
    $\psi \;\;\;\mathrm{theory}$ & $a=-\frac{3}{80}$ & $c=-\frac{1}{120}$ \\    \hline
    $F  \;\;\;\mathrm{theory}$ & $a=\frac{1}{180}$ & $c=\frac{1}{60}$ \\ \hline 
  \end{tabular}
\end{center}

Summing the Weyl anomalies we get the anomaly coefficients of the Liouville SCFT:
\begin{equation}
a = -\frac{3}{16}-Q^2,\quad c = -\frac{1}{8} \ .
\end{equation}

\section{Correlation Functions }
\label{sec:CorrelationFunctions}

In this section we will analyze the correlation functions of vertex operators \eqref{eq:BosonicVertexOp}.
We will relate them  to free fields and four-dimensional Coulomb gas integrals and will
discuss the semiclassical limit and the correlation functions of light primary operators.

\subsection{Relation to Free Fields}
\label{sec:RelationToFreeCS}

Consider the correlation functions of vertex operators  \eqref{eq:BosonicVertexOp}:
\begin{equation}
\label{eq:DefCorrMain}
\mathcal{G}_{\alpha_1\tilde{\alpha}_1,\dots,\alpha_N\tilde{\alpha}_N} (x_1,\dots,x_N) = \left<V_{\alpha_1\tilde{\alpha}_1}(x_1)\cdots V_{\alpha_N\tilde{\alpha}_N}(x_N)\right>,
\end{equation}
where 
\begin{equation} \label{eq:VertexCorrelationFunc}
\left<V_{\alpha_1\tilde{\alpha}_1}(x_1)\cdots V_{\alpha_N\tilde{\alpha}_N}(x_N)\right> \equiv \int D\Phi\, e^{-S_L}\prod_{i=1}^{N} e^{2\alpha_{i}\phi(x_i)+2\tilde{\alpha}_i\phi^{\dagger}(x_i)} \ .
\end{equation}

By shifting $\phi \to \phi-\frac{\log \mu}{3b}$ we get using \eqref{eq:ActionShiftBC} a KPZ scaling relation:
\begin{equation} \label{eq:KPZScale}
\mathcal{G}_{\alpha_1\tilde{\alpha}_1,\dots,\alpha_N\tilde{\alpha}_N} (x_1,\dots,x_N) \propto \mu^{s}\bar{\mu}^{\tilde{s}},~~~~s = \frac{2-2\sum_{i} b\alpha_i}{3b^2},~~~~\tilde{s}= \frac{2-2\sum_{i} b\tilde{\alpha}_i}{3b^2} \ .
\end{equation}
The KPZ scaling relation \eqref{eq:KPZScale} shows that the correlation functions are not analytic in $\mu,\bar{\mu}$ and a naive perturbation theory in $\mu$ 
and $\bar{\mu}$ fails. We denote the chiral and anti-chiral parts of the Liouville interaction terms by:
\begin{equation}
\label{eq:ChiralActions}
\begin{aligned}
S_+ &= \int d^4x\, d^2\theta\, e^{3b\Phi} = \int d^4 x\;3b\left(F-\frac{3}{2}b\psi\psi\right)e^{3b\phi},\\
S_- &= \int d^4x\, d^2\bar{\theta}\, e^{3b\bar{\Phi}} = \int d^4 x\; 3b\left(F^{\dagger}-\frac{3}{2}b\bar{\psi}\bar{\psi}\right)e^{3b\phi^{\dagger}} \ , 
\end{aligned}
\end{equation}
and
\begin{equation}
S_L = S_{C.G.}+\mu S_++\bar{\mu}S_- \ . 
\end{equation}
We separate in the path integral the zero mode $\varphi_0$ of $\mathrm{Re}(\phi)$
and write the decomposition $\phi(x)=\varphi_0+\widehat{\phi}(x)$. Integrating over the zero mode using \eqref{eq:ActionShiftBC} we get:
 \begin{equation}
\begin{aligned}
&\mathcal{G}_{\alpha_1\tilde{\alpha}_1,\dots,\alpha_N\tilde{\alpha}_N} (x_1,\dots,x_N) = \\
&= \int d\varphi_0\, D\widehat{\Phi}\, e^{-S_L\left(\varphi_0+\widehat{\Phi}\right) }\prod_{i=1}^{N} e^{2\alpha_{i}\left(\varphi_0+\widehat{\phi}(x_i)\right)+2\tilde{\alpha}_i\left(\varphi_0+\widehat{\phi}^{\dagger}(x_i)\right)(x_i)}\\
&= \int d\varphi_0\, D\widehat{\Phi}\, e^{-S_{C.G.}\left(\widehat{\Phi}\right)-\frac{4}{b}\varphi_0 -\mu e^{3b\varphi_0}S_+\left(\widehat{\Phi}\right)-\bar{\mu} e^{3b\varphi_0}S_-\left(\widehat{\bar{\Phi}}\right)}\prod_{i=1}^{N} e^{2\alpha_{i}\left(\varphi_0+\widehat{\phi}(x_i)\right)+2\tilde{\alpha}_i\left(\varphi_0+\widehat{\phi}^{\dagger}(x_i)\right)}\\
&= \int D\widehat{\Phi}\, e^{-S_{C.G.}}\prod_{i=1}^{N} e^{2\alpha_{i}\widehat{\phi}(x_i)+2\tilde{\alpha}_i\widehat{\phi}^{\dagger}(x_i)}\int d\varphi_0 \exp\left(-3b(s+\tilde{s})\varphi_0- e^{3b\varphi_0}(\mu S_++\bar{\mu} S_-)\right)\\
&= \int D\widehat{\Phi}\, e^{-S_{C.G}}\prod_{i=1}^{N} e^{2\alpha_{i}\widehat{\phi}(x_i)+2\bar{\alpha}_i\widehat{\phi}^{\dagger}(x_i)} \frac{\Gamma(-s-\tilde{s})}{3b}(\mu S_++\bar{\mu} S_-)^{s+\tilde{s}} \\
&=  \frac{\Gamma(-s-\tilde{s})}{3b}\left<\prod_{i=1}^{N}e^{2\alpha_{i}\widehat{\phi}(x_i)+2\bar{\alpha}_i\widehat{\phi}^{\dagger}(x_i)}(\mu S_++\bar{\mu} S_-)^{s+\tilde{s}}\right>_{C.G.} \ ,
\end{aligned}
\end{equation}
where $\left<\dots\right>_{C.G.}$ denotes expectation values in the free supersymmetric Coulomb gas theory. 

At $s+\tilde{s} = k\in\mathbb{N}\cup\{0\}$ the gamma function diverges, which suggests that the correlation functions \eqref{eq:VertexCorrelationFunc} have poles whenever the sum $s+\tilde{s}$ is a non-negative integer. At these values naive perturbation theory correctly predicts the residues of the correlation function in the variable $2\sum(\alpha_i+\tilde{\alpha}_i)$:
\begin{equation}
\mathcal{G}^{(k)}_{\alpha_1\tilde{\alpha}_1,\dots,\alpha_N\tilde{\alpha}_N} (x_1,\dots,x_N) = \;\underset{2\sum(\alpha_i+\tilde{\alpha}_i)=\frac{4}{b}-3kb}{\mathrm{Res}}\;\mathcal{G}_{\alpha_1\tilde{\alpha}_1,\dots,\alpha_N\tilde{\alpha}_N} (x_1,\dots,x_N) \ .
\end{equation}  
We can then write: 
\begin{equation}
\label{eq:GResidueDecomp}
\begin{aligned}
\mathcal{G}^{(k)}_{\alpha_1\tilde{\alpha}_1,\dots,\alpha_N\tilde{\alpha}_N} (x_1,\dots,x_N)
&=\frac{1}{k!}\left<\prod_{i=1}^{N}V_{\alpha_i\tilde{\alpha}_i}(x_i)(-\mu S_+-\bar{\mu} S_-)^k\right>_{C.G.} \\
&=\sum_{n+\tilde{n} = k} \frac{(-\mu)^n(-\bar{\mu})^{\bar{n}}}{n!\tilde{n}!}\left<\prod_{i=1}^{N}V_{\alpha_i\tilde{\alpha}_i}(x_i)(S_+)^{n}(S_-)^{\tilde{n}}\right>_{C.G.}\\
&\equiv \sum_{n+\tilde{n}=k}(-\mu)^n(-\bar{\mu})^{\tilde{n}}\mathcal{G}^{(n,\tilde{n})}_{\alpha_1\tilde{\alpha}_1,\dots,\alpha_N\tilde{\alpha}_N} (x_1,\dots,x_N) .
\end{aligned}
\end{equation}
From the KPZ scaling relation \eqref{eq:KPZScale} we see that only one term in the sum appearing in equation \eqref{eq:GResidueDecomp} can contribute to the residue. As a result, we can only have a pole when $s,\tilde{s}$ are both non-negative integers which we denote $s=n,\;\tilde{s}=\tilde{n}$. Substituting the definition \eqref{eq:ChiralActions} into equation \eqref{eq:GResidueDecomp} we have:
\begin{equation}
\label{eq:GResiduePsiFDecomp}
\begin{aligned}
&\mathcal{G}^{(n,\tilde{n})}_{\alpha_1\tilde{\alpha}_1,\dots,\alpha_N\tilde{\alpha}_N} (x_1,\dots,x_N) = \\
&= \frac{(3b)^{n+\tilde{n}}}{n!\tilde{n}!} \left<\prod_{i=1}^{N}V_{\alpha_i\tilde{\alpha}_i}(x_i)\prod_{a=1}^{n}\int d^4u_a\left(F-\frac{3b}{2}\psi\psi\right)e^{3b\phi}\prod_{\tilde{a}=1}^{\tilde{n}}\int d^4\tilde{u}_{\tilde{a}}\left(F^{\dagger}-\frac{3b}{2}\bar{\psi}\bar{\psi}\right)e^{3b\phi^{\dagger}}\right>_{C.G.}\\
&= \sum_{\substack{n_F+n_{\psi} = n \\ \tilde{n}_F+\tilde{n}_{\psi} = \tilde{n}}}\frac{(3b)^{n_F+\tilde{n}_F+2n_{\psi}+2\tilde{n}_{\psi}}}{2^{n_{\psi}+\tilde{n}_{\psi}}n_F!\tilde{n}_F!n_{\psi}!\tilde{n}_{\psi}!}
\int d^{4n}u \,d^{4\tilde{n}}\tilde{u}  \left<\prod_{i=1}^{N}V_{\alpha_i\tilde{\alpha}_i}(x_i)\prod_{a=1}^{n}e^{3b\phi(u_a)}\prod_{\tilde{a}=1}^{\tilde{n}}e^{3b\phi^{\dagger}(\tilde{u}_a)}\right>_{C.G.}\\
&\quad~\times\left<\prod_{a=1}^{n_F}F(u_a)\prod_{\tilde{a}=1}^{\tilde{n}_F}F^{\dagger}(\tilde{u}_{\tilde{a}})\right>_{C.G.}\left<\prod_{a=1}^{n_{\psi}}\psi(u_a)\psi(u_a)\prod_{\tilde{a}=1}^{\tilde{n}_{\psi}}\bar{\psi}(\tilde{u}_{\tilde{a}})\bar{\psi}(\tilde{u}_{\tilde{a}})\right>_{C.G.}\\
&\equiv \sum_{\substack{n_F+n_{\psi} = n \\ \tilde{n}_F+\tilde{n}_{\psi} = \tilde{n}}}\frac{(3b)^{n_F+\tilde{n}_F+2n_{\psi}+2\tilde{n}_{\psi}}}{2^{n_{\psi}+\tilde{n}_{\psi}}} \mathcal{G}^{(n_F,\tilde{n}_F,n_{\psi},\tilde{n}_{\psi})}_{\alpha_1\tilde{\alpha}_1,\dots,\alpha_N\tilde{\alpha}_N} (x_1,\dots,x_N) \ ,
\end{aligned}
\end{equation}
where the integration measure is $d^{4n}u\,d^{4\tilde{n}}\tilde{u}\equiv d^4u_1\cdots d^4u_nd^4\tilde{u}_1\cdots d^4\tilde{u}_{\tilde{n}}$.

 In order to compute the residues, which are related to expectation values in supersymmetric Coulomb gas theory, we will use Wick's theorem where the free fields propagators are given in \eqref{eq:FreePropagators}.
 Examining $\mathcal{G}^{(n_F,\tilde{n}_F,n_{\psi},\tilde{n}_{\psi})}_{\alpha_1\tilde{\alpha}_1,\dots,\alpha_N\tilde{\alpha}_N}$ defined in equation \eqref{eq:GResiduePsiFDecomp} we see that in order to get a non-vanishing result we must have $n_F = \tilde{n}_F,n_{\psi} = \tilde{n}_{\psi}$ and therefore $n=\tilde{n}$. 
 
 Consider the three-point function. Conformal invariance implies the general form: 
\begin{equation}
\left<V_{\alpha_1\tilde{\alpha}_1}(x_1)V_{\alpha_2\tilde{\alpha}_2}(x_2)V_{\alpha_3\tilde{\alpha}_3}(x_3)\right> = \frac{C(\alpha_1,\tilde{\alpha}_1,\alpha_2,\tilde{\alpha}_2,\alpha_3,\tilde{\alpha}_3)}{|x_{12}|^{\Delta_1+\Delta_2-\Delta_3}|x_{13}|^{\Delta_1+\Delta_3-\Delta_2}|x_{23}|^{\Delta_2+\Delta_3-\Delta_1}} \ ,
\end{equation}
where $\Delta_i=\Delta_{\alpha_i\tilde{\alpha}_i},\; i=1,2,3$ and the function $C(\alpha_1,\tilde{\alpha}_1,\alpha_2,\tilde{\alpha}_2,\alpha_3,\tilde{\alpha}_3)$ specifies the structure constants of the Liouville SCFT.
 Looking at the residues corresponding to the three-point function and using Wick contraction we get:
\begin{equation}
\begin{aligned}
&\mathcal{G}^{(n_F,n_F,n_{\psi},n_{\psi})}_{\alpha_1\tilde{\alpha}_1,\alpha_2\tilde{\alpha}_2,\alpha_3\tilde{\alpha}_3} (x_1,x_2,x_3) = \frac{\prod_{i<j}|x_{ij}|^{-4\left(\alpha_i\tilde{\alpha}_j+\tilde{\alpha}_i\alpha_j\right)}}{(n_F!)^2(n_{\psi}!)^2}\\
& \times \int d^{4n}u\,d^{4n}\tilde{u} \prod_{a,\tilde{a} \leq n}|u_a-\tilde{u}_{\tilde{a}}|^{-9b^2}\prod_{i=1}^{3}\prod_{a=1}^{n}|x_i-u_a|^{-6b\tilde{\alpha}_i}\prod_{\tilde{a}=1}^{n}|x_i-\tilde{u}_{\tilde{a}}|^{-6b\alpha_i}\\
&\quad~\times\left<\prod_{a=1}^{n_F}F(u_a)\prod_{\tilde{a}=1}^{n_F}F^{\dagger}(\tilde{u}_{\tilde{a}})\right>_{C.G.}\left<\prod_{a=1}^{n_{\psi}}\psi(u_a)\psi(u_a)\prod_{\tilde{a}=1}^{n_{\psi}}\bar{\psi}(\tilde{u}_{\tilde{a}})\bar{\psi}(\tilde{u}_{\tilde{a}})\right>_{C.G.} \ .
\end{aligned}
\end{equation}
In the first non-trivial case $s=\tilde{s}=1$ in the neutral sector $\alpha=\tilde{\alpha}$ we get two integrals, $n_F=\tilde{n}_F=1,\; n_{\psi}=\tilde{n}_{\psi}=0$ and $n_F=\tilde{n}_F=0,\; n_{\psi}=\tilde{n}_{\psi}=1$. The first integral can be evaluated using the formulas derived in \cite{Furlan:2018jlv}. The result is:
\begin{equation}
\begin{aligned}
&\mathcal{G}^{(1,1,0,0)}_{\alpha_1\alpha_1,\alpha_2\alpha_2,\alpha_3\alpha_3} (x_1,x_2,x_3) =\\
&=\prod_{i<j}|x_{ij}|^{-8\alpha_i\alpha_j} \int d^{4}u\,d^{4}\tilde{u}\,|u-\tilde{u}|^{-9b^2}\prod_{i=1}^{3}|x_i-u|^{-6b\alpha_i}|x_i-\tilde{u}|^{-6b\alpha_i}\frac{2}{|u-\tilde{u}|^2}\\
&= 2\pi^2 \prod_{k \neq i<j\neq k}|x_{ij}|^{9b^2-6-8\alpha_i\alpha_j+12b\alpha_k}\frac{\gamma_{2}\left(-\frac{9}{2}b^2-1\right)}{\gamma_{2}\left(-\frac{9}{4}b^2-\frac{1}{2}\right)}\left(\prod_{i=1}^{3}\gamma_2\left(3b\alpha_i\right)\gamma_2\left(3b\alpha_i+\frac{9}{4}b^2+\frac{1}{2}\right)\right)^{-1},
\end{aligned}
\end{equation}
where $\gamma_{2}(x)=\frac{\Gamma(x)}{\Gamma(2-x)}$. The second integral is related to the first by $\mathcal{G}^{(0,0,1,1)}_{\alpha_1\alpha_1,\alpha_2\alpha_2,\alpha_3\alpha_3} (x_1,x_2,x_3) = -2\mathcal{G}^{(1,1,0,0)}_{\alpha_1\alpha_1,\alpha_2\alpha_2,\alpha_3\alpha_3} (x_1,x_2,x_3)$, which results in:
\begin{equation}
\label{eq:CorrFuncExmp}
\begin{aligned}
&\mathcal{G}^{(1,1)}_{\alpha_1\alpha_1,\alpha_2\alpha_2,\alpha_3\alpha_3} (x_1,x_2,x_3) = \prod_{k \neq i<j\neq k}|x_{ij}|^{9b^2-6-8\alpha_i\alpha_j+12b\alpha_k} \\
& \qquad \times 9\pi^2 b^2(2-9b^2)\frac{\gamma_{2}\left(-\frac{9}{2}b^2-1\right)}{\gamma_{2}\left(-\frac{9}{4}b^2-\frac{1}{2}\right)}\left(\prod_{i=1}^{3}\gamma_2\left(3b\alpha_i\right)\gamma_2\left(3b\alpha_i+\frac{9}{4}b^2+\frac{1}{2}\right)\right)^{-1}.
\end{aligned}
\end{equation}

In order to derive the supersymmetric analog of the DOZZ formula we need to perform the integral for general
$n_F$ and $n_{\psi}$, which we leave as an open problem.

\subsection{The Semiclassical Limit}
\label{sec:SemiclassicalLimit}

Consider correlation functions of vertex operators:
\begin{equation}
\left<V_{\alpha_1\tilde{\alpha}_1}(x_1)\cdots V_{\alpha_n\tilde{\alpha}_n}(x_n)\right> \equiv \int D\Phi_c e^{-S_L}\prod_{i=1}^{n} \exp\left(2\left(\frac{\alpha_i}{b}\phi_{c}(x_i)+\frac{\tilde{\alpha}_i}{b}\phi^\dagger_c(x_i)\right)\right) \ ,
\label{integ}
\end{equation}
where we $\Phi_c = b\Phi$. 
We will be interested in evaluating the integral (\ref{integ}) in the semiclassical limit $b\to 0$ using the saddle point approximation. 
The action \eqref{eq:FlatSuperspaceAction} written in terms of $\Phi_c$ scales as $b^{-2}$, thus in order for a vertex operator insertion $V_{\alpha\tilde{\alpha}}$ in \eqref{integ} 
to affect the saddle points we require the scaling  $\alpha,\tilde{\alpha} \sim b^{-1}$. 
In this case we define $\alpha = \eta/b$, $\tilde{\alpha}=\tilde{\eta}/b$ where we keep $\eta$, $\tilde{\eta}$ fixed as $b\to 0$. 
Such vertex operators are called heavy operators and their scaling dimensions \eqref{eq:DimBosonicVertexOp} and R-charges \eqref{eq:U1chargeOfVaa} read:
\begin{equation}
\Delta_{\alpha\tilde{\alpha}}^{\text{heavy}} = \frac{2}{b^2}\left(-2\eta\tilde{\eta}+\eta+\tilde{\eta}\right),~~~~
r_{\alpha\tilde{\alpha}}^{\text{heavy}} = \frac{4}{3b^2}\left(\eta-\tilde{\eta}\right) \ .
\end{equation}
Light operators are  vertex operators with $\alpha = b\sigma$, $\tilde{\alpha}=b\tilde{\sigma}$ where $\sigma$, $\tilde{\sigma}$ are kept fixed as $b\to 0$. Their dimensions and charges in the semiclassical limit read:
\begin{equation}
\Delta_{\alpha\tilde{\alpha}}^{\text{light}} = 2\left(\sigma+\tilde{\sigma}\right) + O(b^2),~~~~
r_{\alpha\tilde{\alpha}}^{\text{light}} = \frac{4}{3}\left(\sigma-\tilde{\sigma}\right)\ .
\end{equation}

Light operators do not affect the saddle point and to lowest order in $b$, the insertion of a light operator results in a multiplication by a $b$-independent factor $e^{2\sigma_i\phi_c(x_i)+2\tilde{\sigma}_i\phi^\dagger_c(x_i)}$ where $\phi_c$ is the saddle point. On the other hand, the insertion of a heavy operator affects the saddle point by adding to the equations of motion \eqref{eq:ComponentEOM} the boundary conditions:
\begin{equation}
\label{eq:HeavyBC}
\phi_c(x) = -2\bar{\eta}\log\left(\left|x-x_i\right|\right)+O(1), \quad \phi_c^{\dagger}(x) = -2\eta\log\left(\left|x-x_i\right|\right)+O(1), \;\quad x\to x_i \ .
\end{equation}
Therefore, the leading exponential asymptotic in the limit $b\to 0$ for the correlation function of heavy and light operators is given by the semiclasssical expression:
\begin{equation} \label{SemiclassCorr}
\begin{aligned}
&\left<V_{\frac{\eta_1}{b}\frac{\tilde{\eta}_1}{b}}(y_1)\cdots V_{\frac{\eta_n}{b}\frac{\tilde{\eta}_n}{b}}(y_n)V_{b\sigma_1 b\tilde{\sigma}_1}(x_1)\cdots V_{b\sigma_m b\tilde{\sigma}_n}(x_m)\right> \\
& \qquad \sim e^{-S_L(\Phi_\eta)}\prod_{i=1}^{m} e^{2\sigma_i\phi_{\eta}(x_i)+2\tilde{\sigma}_i\phi^\dagger_{\eta}(x_i)} \ ,
\end{aligned}
\end{equation}
where $\Phi_{\eta}$ is the solution of the supersymmetric equations of motion obeying the correct boundary conditions \eqref{eq:HeavyBC}. This formula includes effects that are $O(b^{-2})$ in the exponent exactly, while $O(b^0)$ effects are included only if they depend on the positions or conformal dimensions of the light operators. In general there will be more than one solution, and the right hand side will include a sum, or an integral, over the space of solutions.

\subsubsection{Light Primary Operators}

Consider the three-point function of vertex operators of the form \eqref{integ} where all three operators are light. A saddle point solution
is given by \eqref{eq:SuperSol}. In order to calculate the three-point function in the semiclassical approximation 
 one has to integrate over all solutions related to this one by superconformal mappings. Of course, one also needs to show this exhausts the space
 of saddle point solutions.  In appendix \ref{app:2DNequals1SLFT} we show that this scheme indeed works 
for two-dimensional $\mathcal{N}=1$ Liouville SCFT and agrees with the semiclassical limit of the known exact result.
 In \cite{Levy:2018bdc} we used this scheme  and calculated in the saddle point
 approximation the three-point functions of light primary operators of Liouville CFT in an arbitrary
number of  even dimenision. The result agrees with the semiclassical limit of a later exact calculation performed in \cite{Furlan:2018jlv}. 
In both the two-dimensional $\mathcal{N}=1$ Liouville SCFT and the higher dimensional Liouville CFT we use one saddle point solution
and all those related to it by the symmetries of the problem.

All conformal mappings on a domain of $\mathbb{R}^{d}$ for $d>2$ are a composition of translations, dilations, inversions and orthogonal transformation, i.e.\ are higher-dimensional M\"{o}bius transformations. These transformations can be described using $2\times 2$-matrices with entries in the Clifford algebra $C_{d-1} = \mathrm{C}\ell_{0,d-1}(\mathbf{R})$ \cite{Waterman:1993}. In this formalism, higher-dimensional M\"{o}bius transformations can be written as $x \to (\alpha x+\beta)(\gamma x +\delta)^{-1}$ where $\alpha,\beta,\gamma,\delta \in \mathit{\Gamma}_{d-1}\cup \{0\}$, $\alpha\beta^{*},\gamma\delta^{*},\gamma^*\alpha,\delta^*\beta\in\mathbb{R}^{d}$ and $\alpha\delta^*-\beta\gamma^*=1$. This conformal mapping introduces a Weyl transformation with $\sigma = 2\log\left(|\gamma x +\delta|\right)$.  Using this we were able to perform the integral over the symmetry group of the saddle point solution.
In order to perform the integral in our case we need a generalization of the
 the four-dimensional M\"{o}bius transformation to the $\mathcal{N}=1$ superconformal case. This, however,
seem to exists only for even $\mathcal{N}$ \cite{Lukierski:1983jg}. We leave this for further study.

\subsubsection{Heavy Primary Operators}

When considering the opposite case, where all vertex operators are heavy, we are interested only in effects that are $O\left(b^{-2}\right)$ in the exponent. As was done in the two-dimensional case \cite{Hadasz:2007nt} one can integrate out the fermions. This integration is Gaussian, and since in the exponent there is no term that is linear in the fermions, it results in a multiplication by  a functional determinant of an operator involving the three-derivative fermion kinetic operator and the fermions' "masses" (i.e.\ the terms multiplying $\psi\psi$ and $\bar{\psi}\bar{\psi}$ in the action) which are light vertex operators of the form $\mu b^2 e^{3b\phi},\,\bar{\mu} b^2 e^{3b\phi^{\dagger}}$. This functional determinant is of order $O(b^0)$ in the exponent and therefore has no effect on our result. Thus we see that for the calculation of the  correlation function of heavy operators one can set the fermionic fields to zero in the functional integral, which leaves us with an action containing only $\phi,\, F$. 

In the two dimensional case the field $F$ is an auxiliary field with no kinetic term, and therefore it can be integrated out as well by replacing it by its value as dictated by its equations of motion. This reduces the two-dimensional super Liouville action to the two-dimensional ordinary Liouville action and demonstrates that, to leading order, the correlation functions of heavy operators are the same in both theories. In contrast, in the four-dimensional case the field $F$ is a massless propagating field and the interaction term is linear in it. Therefore, trying to integrate it out results as expected in a non-local interaction term of the form $\int e^{3b\phi} \Box^{-1} e^{3b\phi}$ and not in an ordinary Liouville interaction term.
We leave this for a future study.

\section{Summary and Outlook}
 \label{sec:Summary}

We constructed and studied the classical and quantum properties
of an $\mathcal{N}=1$ supersymmetric extension of the four-dimensional Liouville CFT  \cite{Levy:2018bdc}.
There are numerous directions to follow from here.

Solving analytically the integrals for the three-point function of vertex operators is of much importance and it is likely that a DOZZ-like 
formula will be revealed. This can lead to a complete bootstrap solution of the theory.
 Performing the integral over the $\mathcal{N}=1$ superconfromal
group and deriving the three-point function of light primary operators in the semiclassical approximation is quiet valuable as well.
Another interesting direction to follow is the study of deformations of the theory.

As we have seen the $\mathcal{N}=1$ Liouville SCFT in four dimensions exhibits some similarities
to the $\mathcal{N}=2$ Liouville SCFT in two dimensions such as a non-renormalization of the classical background charge.
It would be interesting to see if there is also an $\mathcal{N}=1$ four-dimensional duality relation reminiscent of the one between
the $\mathcal{N}=2$ Liouville SCFT and the two-dimensional fermionic black hole \cite{Hori:2001ax}. 
From \eqref{eq:DimBosonicVertexOp} and \eqref{eq:U1chargeOfVaa} one sees  that the vertex  operator $V_{\frac{1}{b}}= e^{\frac{2}{b}(\phi+\phi^{\dagger})}$ has zero scaling dimension and R-charge. $\mathcal{N}=2$ Liouville SCFT in two dimensions exhibits a similar phenomena and it has been used
in \cite{Ahn:2002sx,Nakayama:2004vk}  to argue for the existence of a dual decription of the theory. This structure and its interpretation
deserves a further study both in two and four dimensions.

One can generalize the $\mathcal{N}=1$ Liouville SCFT and construct an  $\mathcal{N}=2$ Liouville SCFT in four dimensions where the quaternionic formalism \cite{Lukierski:1983jg} may be useful for solving the model. One can also go to higher dimensions and construct six-dimensional 
$\mathcal{N}=1$ and   $\mathcal{N}=2$ Liouville SCFTs
by generalizing the concepts of supersymmetric Weyl covariant differential operator and supersymmetric  ${\cal Q}$-curvature.

The study of the higher-dimensional Liouville SCFTs in the presence of boundaries is likely to reveal a rich structure.
Finally one can consider odd dimensions where Liouville SCFTs are non-local
with a pseudo-diffrenetial  Weyl covariant operator and a non-local ${\cal Q}$-curvature.

\vskip 1cm
{\bf \large Acknowledgment } 
\vskip 0.5cm
We would like to thank Y. Nakayama for a discussion. This work is supported in part by the I-CORE program
of Planning and Budgeting Committee (grant number 1937/12), the US-Israel Binational Science Foundation, GIF and the ISF Center of Excellence. T.L gratefully acknowledges the support of the Alexander Zaks Scholarship. A.R.M gratefully acknowledges the support of the Adams Fellowship Program of the Israel Academy of Sciences and Humanities.

\appendix 

\section{Notations and Conventions}
\label{app:SUGRANotations}
We use the notations of \cite{Wess:1992cp} in Lorentzian signature. These were adapted to Euclidean signature in appendix A of \cite{Dumitrescu:2012ha}.
We denote the conjugate of the spinor field $\psi_\alpha$ by $\bar{\psi}_{\dot{\alpha}} = (\psi_\alpha)^\dagger$, where $\lbrace \alpha, \dot{\alpha} \rbrace = \lbrace 1,2 \rbrace$.

The $\mathcal{N}=1$ supergravity multiplet includes the vielbein $e_m^a(x)$, the gravitino $\psi_m^\alpha(x),\bar{\psi}_m^{\dot{\alpha(x)}}$ and the auxiliary fields $M$ and $b_a$. 
It can be formulated in terms of three superfields: $R$, $W_{\alpha \beta \gamma}$ and $G_a$.
$G_a$ is a real superfield $G_{\alpha\dot{\alpha}}^\dagger =G_{\alpha\dot{\alpha}}$,   $R$ and  $W_{\alpha \beta \gamma}$ are chiral  superfields:
$\mathcal{\bar{D}}_{\dot{\alpha}} R = \mathcal{\bar{D}}_{\dot{\alpha}}W_{\alpha \beta \gamma} = 0$
where $\mathcal{D}$ is the covariant superderivative. $W_{\alpha \beta \gamma}$ is completely symmetric in all three indices.  
Their anti-chiral complex conjugated partners are denoted by $\bar{R}$ and $\bar{W}_{\alpha \beta \gamma}$, respectively. 

One has \cite{Stelle:1978ye,Ferrara:1978em,Wess:1992cp}: 
\begin{align}
&R\mid_{\theta = \bar{\theta} = 0} = -\frac{1}{6}M, \nonumber \\
&G_a \mid_{\theta = \bar{\theta} = 0} = -\frac{1}{3}b_a \ .
\label{AUX}
\end{align}
In flat space a chiral superfield $\bar{D}_{\dot{\alpha}}\Phi = 0$ can be expanded as:
\begin{equation}
\Phi = \phi (x_+) +\sqrt{2} \theta \psi(x_+) +\theta \theta F(x_+),~~~~x_+^m = x^m + i \theta \sigma^m \bar{\theta} \ ,
\end{equation}
where
\begin{equation}
\phi = \Phi \mid_{\theta = \bar{\theta}=0}, \qquad \psi_\alpha = \frac{1}{\sqrt{2}}D_\alpha\Phi \mid_{\theta=\bar{\theta}=0}, \qquad F = -\frac{1}{4}D^2 \Phi \mid_{\theta=\bar{\theta}=0},
\end{equation}
One defines new $\Theta_\alpha$ variables such that the expansion coefficients of the chiral superfields are the covariant components: 
\begin{equation}
\Phi = \phi(x) + \sqrt{2}\Theta^\alpha\psi_\alpha(x) + \Theta\Theta F(x),
\end{equation}
and the $\Theta$ variables carry local Lorentz indices.  
One has:
\begin{equation}
\begin{aligned}
R &= -\frac{1}{6} \left( M +\Theta\left( \sigma^a\bar{\sigma}^b\psi_{ab} - i\sigma^a\bar{\psi}_a M +i\psi_a b^a\right) \right.\\
& \qquad +\Theta\Theta \left( -\frac{1}{2}\mathcal{R}+i\bar{\psi}^a\bar{\sigma}^b\psi_{ab} +\frac{2}{3}MM^* \right. \\
& \qquad \qquad \qquad \left.+\frac{1}{3}b^a b_a-ie_a^m\mathcal{D}_m b^a +\frac{1}{2}\bar{\psi}\bar{\psi}M -\frac{1}{2}\psi_a \sigma^a\bar{\psi}_c b^c \right.\\
& \qquad \qquad \qquad \left. \left. +\frac{1}{8}\epsilon^{abcd}\left( \bar{\psi}_a\bar{\sigma}_b \psi_{cd} + \psi_a\sigma_b\bar{\psi}_{cd}\right) \right) \right),
\end{aligned}
\end{equation}
and $\mathcal{R}$ denotes the bosonic Ricci scalar. 

The chiral superspace integral is
invariant under supergravity trnasformation: 
\begin{equation}
\int d^2 \Theta\, 2\mathcal{E} W(\Phi),
\end{equation}  
where $W(\Phi)$ stands for any holomorphic function of $\Phi$. $\mathcal{E}$ is the chiral density: 
\begin{equation}
\mathcal{E}=\frac{e}{2}\left( 1 + i \Theta \sigma^a\bar{\psi}_a -\Theta\Theta\left(M^*+\bar{\psi}_a\bar{\sigma}^{ab}\bar{\psi}_b \right) \right) \ ,
\label{EPS}
\end{equation}
and  $e = \text{det}(e_m^a)$.

\section{Light Operators Three-Point Function in Two-Dimensional $\mathcal{N}=1$  Liouville SCFT}

\label{app:2DNequals1SLFT}
The two-dimensional $\mathcal{N}=1$ Liouville SCFT action reads \cite{Polyakov:1981rd,Distler:1989nt}:
\begin{equation}
\begin{aligned}
S &= \frac{1}{2\pi}\int d^2z\,d\theta\,d\bar{\theta} \left(D\Phi\bar{D}\Phi +4\pi i\mu e^{b\Phi}\right)\\
&= \frac{1}{2\pi}\int d^2z \left(\partial\phi\bar{\partial}\phi+\psi\bar{\partial}\psi+\bar{\psi}\partial\bar{\psi}-F^2+4\pi\mu\left(ib^2\psi\bar{\psi}-bF\right)e^{b\phi}\right),
\end{aligned}
\end{equation}
where $\phi$ is a real Liouville field, $F$ is a real bosonic auxiliary field and $\psi$ is a fermionic field with two off-shell degrees of freedom. The dimensionless parameters are the cosmological constant $\mu$, the background charge $Q$ and $b$. The Liouville field $\phi$ has background charge $Q$, i.e.\ it satisfies the boundary conditions:
\begin{equation}
\phi = -Q\log(z\bar{z})+O(1),\quad |z|\to \infty  \ .
\label{BC}
\end{equation}
The classical field equations read:
\begin{align}
\partial\bar{\partial}\phi &= 2\pi\mu b^2\left(ib\psi\bar{\psi}-F\right)e^{b\phi}, \nonumber\\
\bar{\partial}\psi &= 2\pi i \mu b^2\bar{\psi}e^{b\phi}, \nonumber\\
F &= -2\pi\mu b e^{b\phi} \ .
\end{align}
We will work in the semiclassical limit $b\to 0$  and calculate the three-point function of light vertex operators in 
 the saddle point approximation. We define $\Phi_c = b\Phi$ and look for saddle points of the action. There are no real solutions of the field
 equations  that satisfy the boundary conditions \eqref{BC}, and we will look for complex saddle points. Our starting point is the sphere solution, which when analytically continued corresponds to the supergravity multiplet of the supersymmetric sphere of radius $r$:
\begin{equation}
\label{eq:2dClassicalSolEx}
\phi_c = -\log\left(\frac{|z|^2+1}{2r}\right), \qquad \psi_c = 0,\qquad F_c= \frac{i}{|z|^2+1}, 
\end{equation}
with $r = \pm \frac{i}{4\pi\mu b^2}$ imaginary. Without loss of generality we choose $r=-\frac{i}{4\pi \mu b^2}$ so that $\mu b = -\frac{i}{4\pi r b}$. 

In terms of the superfield the classical solution \eqref{eq:2dClassicalSolEx} is given by 
\begin{equation}
\label{eq:ClassicalSuperfield2DSim}
\Phi_c = \phi_c +i \theta\bar{\theta}F_c.
\end{equation}
Since the theory is superconformally invariant,  in order to use the saddle point approximation we have to integrate over all solutions related to \eqref{eq:ClassicalSuperfield2DSim} by a superconformal transformation. All superconformal transformations can be written as super-M\"{o}bius transformations which read \cite{Gieres:1992sc}:
\begin{equation}
z' = \frac{az+b}{cz+d}+\theta\frac{\alpha z+\beta}{(cz+d)^2},~~~~\theta' = \frac{\alpha z+ \beta}{cz+d}+  \frac{\theta}{cz+d} \ ,
\end{equation}
where $a,b,c,d$ are complex numbers, $\alpha, \beta$ are complex Grassmann numbers and  $ad-bc = 1+\alpha\beta$. The supergroup of super-M\"{o}bius transformations is isomorphic to the Lie supergroup $\mathrm{OSp}(2;1)$ which has the supermatrix representation:
\begin{equation}
\mathrm{OSp}(2;1) = \left\lbrace \left. \left( \begin{array}{ccc}
a & b & b\alpha-a\beta \\ 
c & d & d\alpha-c\beta \\ 
\alpha & \beta & 1-\alpha\beta
\end{array} \right) \right|\; a,b,c,d \in \mathbb{C},\; ad-bc =1 +\alpha\beta \right\rbrace \ .
 \end{equation} 
 The superdeterminants of all supermatrices $M\in\mathrm{OSp}(2;1)$ are equal to one, $\mathrm{sdet}(M)=1$.
 
 Under super-M\"{o}bius transformation the Liouville superfield solution \eqref{eq:ClassicalSuperfield2DSim} transforms as:
\begin{equation}
\label{eq:2dLiouvilleTransBasic}
\Phi_c' = \phi_c\left(z',\bar{z}'\right)+i\theta'\bar{\theta}'F_c(z',\bar{z}')-\sigma \ ,
\end{equation}
where $\sigma$ is the super-Weyl factor associated with super-M\"{o}bius transformation:
\begin{equation}
\sigma=\log\left(\left|cz+d+\delta\theta\right|^2\right),
\end{equation}
and $\delta = \alpha d +\beta c$.

A straightforward calculation yields the following expression for the zeroth and second order components in the $\theta$ expansion after a super-M\"{o}bius transformation:
\begin{eqnarray}
\label{eq:ScalarAfterSuperMob}
\Phi_c' \mid_{\theta=\bar{\theta}=0} &=& \log(2r)-\log\left(|az+b|^2+|cz+d|^2\right)-\frac{|\alpha z+\beta|^2}{|az+b|^2+|cz+d|^2}, \nonumber\\
\Phi_c' \mid_{\theta\bar{\theta}}&=&i\theta\bar{\theta}\left(\frac{i}{|az+b|^2+|cz+d|^2}-\frac{i|\alpha z+\beta|^2}{\left(|az+b|^2+|cz+d|^2\right)^2}\right) \ .
\end{eqnarray}
The fermionic parts were ignored for simplicity. One can verify that \eqref{eq:ScalarAfterSuperMob} satisfy the parts of the equations of motion that are independent of the fermions (the equation of motion for the auxiliary field and the zeroth-order term of the scalar equation of motion in the Grassmann numbers).   
We use the zeroth component of this result to write light vertex operators evaluated at the saddle point:
\begin{equation}
\begin{aligned}
e^{\sigma_i\phi_c} &= (2r)^{\sigma_i}\frac{1}{\left(|az+b|^2+|cz+d|^2\right)^{\sigma_i}}\left(1-\sigma_i\frac{|\alpha z+\beta|^2}{|az+b|^2+|cz+d|^2}\right) \ .
\end{aligned}
\end{equation}
Using the conformal invariance of the correlation function, we set our three light vertex operators at points $z= 0,1,\infty$ and write the structure constants as:
\begin{equation}
\begin{aligned}
\label{eq:Calc12dCorr}
C(b\sigma_1,b\sigma_2,b\sigma_3) & = \left<V_{b\sigma_1}(0)V_{b\sigma_2}(1)V_{b\sigma_3}(\infty)\right> \\
& \sim \int d\mu(a,b,c,d,\alpha,\beta) e^{\sigma_1\phi(0)}e^{\sigma_2\phi(1)}e^{\sigma_3\phi(\infty)} \ ,
\end{aligned}
\end{equation}
where $d\mu(a,b,c,d,\alpha,\beta)$ is a super-Haar measure on the supergroup $\mathrm{OSp}(2;1)$:
\begin{equation}
d\mu(a,b,c,d,\alpha,\beta) = \delta^2(ad-bc-1-\alpha\beta)d^2a\,d^2b\,d^2c\,d^2d\,d^2\alpha\,d^2\beta.
\end{equation}
Equation \eqref{eq:Calc12dCorr} yields:
\begin{equation}
\begin{aligned}
\label{eq:Calc22dCorr}
&\int d\mu(a,b,c,d,\alpha,\beta) \left( \frac{(2r)^{\sigma_1}}{\Sigma_1^{\sigma_1}}\left( 1-\sigma_1 \frac{|\beta|^2}{\Sigma_1}\right)\frac{(2r)^{\sigma_2}}{\Sigma_2^{\sigma_2}}\left(1-\sigma_2\frac{|\alpha+\beta|^2}{\Sigma_2}\right) \right.\\
& \qquad \qquad\qquad \qquad \qquad \left. \frac{(2r)^{\sigma_3}}{\Sigma_3^{\sigma_3}}\left(1-\sigma_3\frac{|\alpha|^2}{\Sigma_3}\right) \right),
\end{aligned}
\end{equation}
where $\Sigma_1 = |b|^2+|d|^2$, $\Sigma_2 = |a+b|^2+|c+d|^2$ and $\Sigma_3 = |a|^2+|c|^2$. Equation \eqref{eq:Calc22dCorr} reads:
\begin{equation}
\begin{aligned}
\label{eq:Calc32dCorr}
(2r)^{\sigma_1+\sigma_2+\sigma_3}\int & d\mu(a,b,c,d,\alpha,\beta)  \frac{1}{\Sigma_1^{\sigma_1}\Sigma_2^{\sigma_2}\Sigma_3^{\sigma_3}}\left(1-\frac{\sigma_1}{\Sigma_1}|\beta|^2  -\frac{\sigma_2}{\Sigma_2}|\alpha+\beta|^2 -\frac{\sigma_3}{\Sigma_3}|\beta|^2  \right.\\
& \qquad \qquad \qquad \left. + \frac{\sigma_1\sigma_2}{\Sigma_1\Sigma_2}|\alpha|^2|\beta|^2+\frac{\sigma_1\sigma_3}{\Sigma_1\Sigma_3}|\alpha|^2|\beta|^2 +\frac{\sigma_2\sigma_3}{\Sigma_2\Sigma_3}|\alpha|^2|\beta|^2\right).
\end{aligned}
\end{equation}
With a change of variables $x \to (1+\frac{1}{2}\alpha\beta)x$, for $x=\lbrace a,b,c,d \rbrace$,  \eqref{eq:Calc32dCorr} yields: 
\begin{equation}
\begin{aligned}
\label{eq:Calc42dCorr}
(2r)^{\sigma_1+\sigma_2+\sigma_3}\int & d\mu_B  \frac{1}{\Sigma_1^{\sigma_1}\Sigma_2^{\sigma_2}\Sigma_3^{\sigma_3}} \int d^2 \alpha d^2 \beta \left\vert 1+\left(1-\frac{\sigma_1+\sigma_2+\sigma_3}{2}\alpha\beta\right) \right\vert^2  \\
& \left(1-\frac{\sigma_1}{\Sigma_1}|\beta|^2  -\frac{\sigma_2}{\Sigma_2}|\alpha+\beta|^2 -\frac{\sigma_3}{\Sigma_3}|\beta|^2 \right.\\
& \qquad \left.  + \left(\frac{\sigma_1\sigma_2}{\Sigma_1\Sigma_2}+\frac{\sigma_1\sigma_3}{\Sigma_1\Sigma_3} +\frac{\sigma_2\sigma_3}{\Sigma_2\Sigma_3}\right)|\alpha|^2|\beta|^2\right),
\end{aligned}
\end{equation}
where $d\mu_B \equiv \delta^2(ad-bc-1) d^2a d^2b d^2cd^2d$ is the bosonic measure. Integration over the grassman variables yields: 
\begin{equation}
\begin{aligned}
\label{eq:Calc52dCorr}
(2r)^{\sigma_1+\sigma_2+\sigma_3}\int & d\mu_B \left( -\frac{1}{\Sigma_1^{\sigma_1}\Sigma_2^{\sigma_2}\Sigma_3^{\sigma_3}}\left(1-\frac{\sigma_1+\sigma_2+\sigma_3}{2}\right)^2 \right. \\
& \left. + \frac{\sigma_1\sigma_2}{\Sigma_1^{\sigma_1+1}\Sigma_2^{\sigma_2+1}\Sigma_3^{\sigma_3}}+\frac{\sigma_1\sigma_3}{\Sigma_1^{\sigma_1+1}\Sigma_2^{\sigma_2}\Sigma_3^{\sigma_3+1}}+\frac{\sigma_2\sigma_3}{\Sigma_1^{\sigma_1}\Sigma_2^{\sigma_2+1}\Sigma_3^{\sigma_3+1}} \right) \ .
\end{aligned}
\end{equation}
From \cite{Zamolodchikov:1995aa}:
\begin{equation}
\int \frac{d\mu_B}{\Sigma_1^{\sigma_1}\Sigma_2^{\sigma_2}\Sigma_3^{\sigma_3}} = \frac{\Gamma\left(\frac{\sigma_{1+2+3}}{2}\right)\Gamma\left(\frac{\sigma_{1+2-3}}{2}\right)\Gamma\left(\frac{\sigma_{1-2+3}}{2}\right)\Gamma\left(\frac{\sigma_{-1+2+3}}{2}\right)}{\Gamma(\sigma_1)\Gamma(\sigma_2)\Gamma(\sigma_3)} \ ,
\end{equation}
where $\sigma_{i+j \pm k} \equiv \sigma_i+\sigma_j \pm \sigma_k$, and $\lbrace i,j,k \rbrace = \lbrace 1,2,3\rbrace$. Using $\frac{x}{\Gamma(x+1)} = \frac{1}{\Gamma(x)}$, the integral \eqref{eq:Calc52dCorr} yields: 
\begin{equation}
(2r)^{\sigma_1+\sigma_2+\sigma_3}\frac{\Gamma\left(\frac{\sigma_1+\sigma_2+\sigma_3}{2}\right)\Gamma\left(\frac{\sigma_1+\sigma_2-\sigma_3}{2}\right)\Gamma\left(\frac{\sigma_1-\sigma_2+\sigma_3}{2}\right)\Gamma\left(\frac{-\sigma_1+\sigma_2+\sigma_3}{2}\right)}{\Gamma(\sigma_1)\Gamma(\sigma_2)\Gamma(\sigma_3)} \ ,
\end{equation}
which agrees perfectly with the $b \to 0$ limit of the exact result (see e.g.\ \cite{Fukuda:2002bv,Rashkov:1996np,Nakayama:2004vk}).

\end{document}